# Returning Scientists and the Emergence of China's Science System

Cong Cao[1], Jeroen Baas[2], Caroline S. Wagner[3], Koen Jonkers[4]*

[1]Faculty of Business, The University of Nottingham Ningbo China, 199 Taikang East Road, Ningbo 315100 China

[2]Elsevier B.V. Registered Office: Radarweg 29, 1043 NX Amsterdam, The Netherlands

[3] John Glenn College of Public Affairs, The Ohio State University, Columbus, OH, USA 43210

[4]European Commission, Joint Research Centre, Avenue Champs de Mars 21, Brussels, Belgium

*correspondence: Koen Jonkers, European Commission, Joint Research Centre, Avenue Champs de Mars 21, Brussels, Belgium, koen.jonkers@ec.europa.eu

Abstract

China's approach to developing a world-class science system includes a vigorous set of programmes to attract back Chinese researchers who have overseas training and work experience. No analysis is available to show the performance of these mobile researchers. This article attempts to close part of this gap. Using a novel bibliometric approach, we estimate the stocks of overseas Chinese and returnees from the perspective of their publication activities, albeit with some limitations. We show that the share of overseas Chinese scientists in the US is considerably larger than that in the EU. We also show that Chinese returnees publish higher impact work, and continue to publish more and at the international level than domestic counterparts. Returnees not only tend to publish more, but they are instrumental in linking China into the global network. Indeed, returnees actively co-publish with researchers in their former host system, showing the importance of scientific social capital. Future research will examine the impact of length of stay, among other factors, on such impact and integration.

Disclaimer: This paper and findings presented in this paper do not necessarily represent the views of the European Commission. The European Commission nor anyone acting on its behalf can be held responsible for the use thereof. Jeroen Baas is an Elsevier employee. Elsevier runs Scopus, which is the source of the data used to approximate mobility and mobility impact.
Funding acknowledgement: Cong Cao would like to acknowledge the use of funding from the European Commission and the National Natural Science Foundation of China grant # 71774091.
Acknowledgements: comments from Daniel Vertesy, Thomas Zacharewicz, Florian Flachenecker, Alice Szczepanikova, Peter Fako and anonymous referees are gratefully acknowledged – the usual disclaimer applies.

1. Introduction

For the past 40 years, China has made investments in building domestic research capacity, in prioritizing advancements in targeted areas of science and technology, and in seeking opportunities of attracting foreign investment and knowledge people.  Against such backdrop and the unprecedented entrance of a huge nation into the global science system, this study examines the rise in number of Chinese scientific personnel, their international engagement, the number of overseas Chinese scientists and returnees, and the contribution of such a pool of talent to Chinese and to global science.

This article discusses and analyses China's human resources development amid its policy-driven process of globalization and internationalization.  The paper starts with a discussion of human capital development in the Chinese research and innovation system. This is followed by an examination of Chinese overseas study in the reform and open-door era, and especially the cross-border mobility of highly skilled Chinese talent. China has implemented the largest and most intense campaign to attract overseas Chinese scientists to return. Due to a lack of reliable statistics, it is difficult to estimate the exact number of overseas Chinese scientists and engineers and returnees (especially from the EU). Rather than assessing the success of any specific return migration programme, this paper extracts data from S&T publications to offer a conservative estimate of the number of Chinese scientists in Europe and in the United States, as well as estimate of the number of returnees to China from these countries. These estimates are based on a novel bibliometric approach explained in the methodological section. The paper proceeds with an assessment of the role which these overseas Chinese scientists and returnees have played in the formation of transnational ties among China, Europe and United States and thus in embedding the Chinese science system in transnational collaborative networks.

We expect to show that increasing numbers of Chinese researchers have returned home over time, that those who return home are more likely to contribute to higher impact science and that they are more prone to engage in international collaboration. These international collaboration patterns are furthermore expected to be shaped by past mobility experience: i.e. we expect to be able to measure the effect of scientific social capital aspect of scientific and technological human capital (Jonkers & Tijssen, 2008; Jonkers & Cruz-Castro, 2013; Bozeman, Dietz & Gaughan, 2001).

2. Growth of Chinese Science

In the span of four decades, China has evolved from a peripheral player to become one of the world's most productive science systems (Zhou & Leydesdorff, 2006). A variety of indicators suggest that China's science and technology (S&T) capabilities are on a sharply rising trajectory.  The Organization for Economic Cooperation and Development (OECD, 2010) reports that since the 1990s, spending on S&T and research and development (R&D) in China has been increasing at a rate faster than that of overall economic growth.  In 2017,

China reported to spend RMB1.75 trillion ($259 billion), or 2.12% of its increasing gross domestic product (GDP), on R&D (NBS, 2018). This is higher than that of European Union, which, with 317 billion euro ($358 billion), arrives at 2.06% of GDP, while still lower than that of the United States, whose respective figures were estimated at $510 billion and 2.74% of GDP in 2016 (Van Noorden, 2014; NSF, 2018; ESTAT, 2018). China's exponential rise in the S&T terms can be attributed to reform and open-door policies, especially the S&T system since the mid-1980s (Appelbaum et al., 2018). The leadership's vision on the role of S&T in the country's modernization and government's commitment in S&T also are contributing factors, evidenced in increasing public expenditure in S&T and R&D.

In the recent decade, there has been a steady rise in the contributions of Chinese scientists to international publications (NSF, 2018). Measured by the number of papers published in journals indexed by *Scopus*, China ranked second in the world in number of publications, accounting for some 19% (18% fractional) of the world's total in 2017, up from 10% (9% fractional) in 2005 (see also Leydesdorff et al, 2018). Leydesdorff et al. (2014) showed that the yearly growth for China's in the top 1% of highly cited articles was 0.85 percentage points per year between 2000 and 2012. In 2005, China contributed 5% (4% fractional) of the top 10% most highly cited papers (field normalised); but in 2017, that statistic rose to 9% (8% fractional: that is fractionalised and field normalised). That is, based on this measure China is still somewhat below the world average of 10% in terms of high impact publications as share of its total output, but it has shown a remarkable increase in the share of impactful papers while expanding its total publication output exponentially.[i] (FWCI stands for Field Weighted Citation Impact, a widely accepted method of normalizing citation measures across fields (Purkayastha, Palmaro, Falk-Krzesinski, & Baas, 2019)). Taking the non-field normalised measure of the share of top 10% most highly cited papers, the rise is even more spectacular: tripling rather than doubling between 2005 and 2017. The reason why China's share of field weighted top 10 % papers grows slower than the absolute number is that in chemistry and engineering, fields in which China is heavily specialized, citations receive a lower weight when normalized by subject field.

**Table 1 Bibliometric indicators of China's scientific performance**

| Year | PP10 FWCI[1] FULL | PP10 FWCI FRAC | PP10 CITS FULL | PP10 CITS FRAC | WLD share full | WLD share frac | intl share of CHN pubs full | intl share of CHN pubs frac |
|---|---|---|---|---|---|---|---|---|
| 2005 | 5% | 4% | 5% | 4% | 10% | 9% | 14% | 8% |
| 2010 | 6% | 5% | 7% | 6% | 16% | 15% | 14% | 8% |
| 2017 | 9% | 8% | 15% | 13% | 19% | 18% | 22% | 14% |

---
[1]

International co-publications also have played a considerable role in the rise in impact. An increasing share – from 14% (8% fractional) in 2005 to 22% (14% fractional) in 2017 – of the rapidly rising number of Chinese publications are the result of international co-publications. China's international co-publications with the EU and US on average receive considerably more citations than China's articles more generally. One also observes that international co-publications with China receive more citations on average than the average publication from the EU28 and the US (see also Leydesdorff, et al., 2014). The rapid development of China's international collaboration is thus mutually beneficial for both China and its partners (Adams, 2012). Numbers of co-publications between China and the US have outpaced EU-US co-publications, while EU-China co-publications have grown less rapidly than EU-US co-publications (see also Wagner et al, 2015; Preziosi et al, 2019).

**Figure 1 Trends in international co-publications**

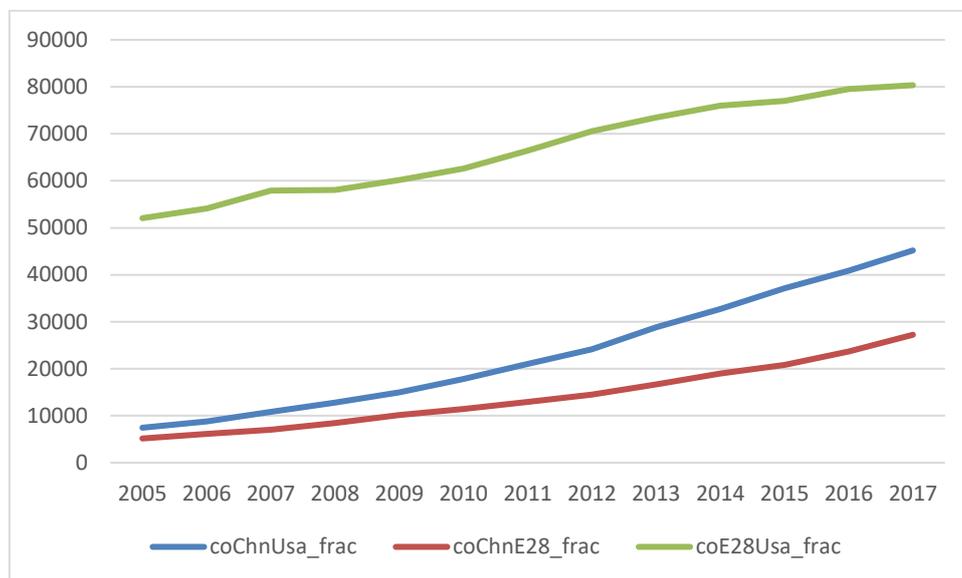

| Year | Chn-Usa | Chn-E28 | E28-USA |
|------|---------|---------|---------|
| 2005 | 7,452.1 | 5,154.6 | 52,048.7 |
| 2010 | 17,823.3 | 11,434.4 | 62,600.0 |
| 2017 | 45,181.3 | 27,231.3 | 80,352.3 |

Funding intensity and the level of institutional for research systems affect the scientific performance of research systems (e.g. Sandström and van den Besselaar, 2018; Cimini et al, 2016); the performance is also influenced by the presence of high-quality scientific human capital, scientific mobility, and international research collaboration (Wagner and Jonkers, 2017; Sugimoto et al, 2017). Scientific mobility can be characterised "as a way to augment the scientist's professional network and the resources available, thus increasing the scientist's scientific and technical human capital" (Edler et al, 2011, p. 792).

The literature has suggested theoretically (e.g; Borjas, 1994) and shown empirically (e.g., Gaule and Piacentini, 2013) that people are positively selected into migration: the most talented obtain education abroad or get job offers abroad. It is therefore expected that returnees outperform natives who did not move. Depending on the circumstances, the choice to return may also be subjected to selection (Borjas and Bratsberg, 1996). Returnees may be either positively selected: i.e., the best of the best return; or negatively selected: i.e., the worst of the best return. Overall, if scientific potential would be an innate – non-context dependent – quality, both groups of returnees may be expected to outperform the natives that never migrated.

Smaller scale studies suggest that returned scientists indeed publish higher impact papers than those domestic counterparts without foreign work experience (Jonkers and Tijssen, 2008; Baruffaldi and Landoni, 2012; Jonkers and Cruz-Castro, 2013). Gibson and McKenzy (2014) do not find the same effect on high impact publications (for small island countries), but do share the aforementioned authors' findings on the greater propensity of returnees for international collaboration. Van Holm et al. (2019) found that Chinese scientists in the US maintain more robust and productive networks than their US born counterparts. Diaspora scientists tend to cooperate to a relatively large extent with those in the home system (Jin et al, 2007; Wagner, 2009; Scellato et al, 2015) either spontaneously or in response to scientific diaspora policies implemented by the government of their home system (Jonkers, 2008; Meyer, 2001). Through this channel, they may thus help in the further development of their home research system (Agrawal et al., 2011; Wagner et al., 2018).

### 3. Public Policy Changes to Improve the Quality of China's Scientific workforce

Underlying the impressive performance taking place in the Chinese S&T system is the emergence of a very large talent pool whose quality has been improving. As a concept, talent (*rencai*) has gained increasing popularity and significance in China since the turn of the twenty-first century, when the leadership realized that "empowering the nation with talent" (*rencai qiangguo*) is key to "rejuvenating the nation with science, technology, and education" (*kejiao xingguo*), a strategy introduced in the mid-1990s. Despite possessing the world's largest number of scientists and engineers and having a very full pipeline in higher education, China has been facing a serious talent challenge, especially at the high end. Critical concerns proliferated from the political and scientific leadership to enterprise chief executives, including the country heads of multinational corporations operating in China, about whether China's potential could be realized given the uncertainties surrounding the demand and supply, quantity and quality, and effective utilization of China's current and future S&T workforce (Simon & Cao, 2009).

In 2006, China launched the Medium and Long-Term Plan for the Development of Science and Technology (2006–2020) (MLP), signalling the nation's commitment to rely more on "brain" than "brawn" to bring China into a strong, leadership position (Cao, Suttmeier & Simon, 2006). The Central Leading Group for Coordinating Talent Work (CLGCTW), a high level task force newly established by the Central Committee of the Chinese Communist Party (CCPCC) within its organisation department, led the implementation of the Plan on National Medium and Long-Term for the Development of Talent (2010–2020).  The formulation of the plan highlighted the urgency that China placed on achieving five goals: 1) transforming China's population dividend into a talent dividend; 2) pursuing a shift from a "made in China" to a "created in China" model; 3) focusing less on attracting foreign capital and more on attracting human capital; 4) emphasizing the importance of "software" rather than "hardware"; and 5) shifting from an investment model to an innovation model.  In addition to providing guiding policies and strategic goals, the plan recommended national talent development targets; specified sectors in which talent is in great demand; called for establishing national programs to support and nurture the development of talent in various fields; and prioritized areas in which improvements in policy and institution-building are necessary to better employ talent (CCPCC, 2010; Wang, 2011; Simon & Cao, 2011).

In 2015, The CCPCC and China's State Council released the innovation-driven development strategy. In 2016, the CCPCC issued the "Opinions on Deepening the Reform of the Institutional Mechanism for Talent Development" to accelerate the talent-driven nation building, stimulate fully Chinese people's innovativeness, creativity, and entrepreneurship, and attract talent from all walks of life to fulfil the cause of the party-state.[ii]

These policy actions were the culmination of those in the 1990s when the Chinese government started making large investments in the reform and upgrading of the country's universities through its 211 and 985 Programmes (Zhang, Patton & Kenney, 2012; Simon & Cao, 2009; Wang, Wang & Liu, 2011). In 2017, these higher education programmes were merged to become the "Double First-Class University Programmes" aiming to raise rankings of 30 Chinese universities to the list of the top 100 leading universities worldwide (Qi Wang, 2017; Zhang, 2018). The Chinese Academy of Sciences, the country's leading research institution, also experienced large scale reform and investments (Suttmeier, Cao & Simon, 2008; Cao and Suttmeier, 2017). As a consequence China has not only been able to train increasingly larger volumes of scientists, but these investments and reforms have also made China a more attractive point of return for overseas Chinese scientists.

Domestically, between 2000 and 2016, the total number of Chinese postgraduate students with a master's degree increased by tenfold while the percentage of those in science, engineering, agriculture, and medicine (SEAM[iii]) showed a slight decline in the early twenty-first century before stabilizing in recent years at 57% of all students. Similarly, entering the twenty-first century, there was an initial decline in the number of doctoral students majoring in SEAM; but the trend was reversed.  In 2016, 77% of Chinese doctorates were awarded in SEAM. The annual number of SEAM PhDs graduating from Chinese universities has grown by 450% since 2000. At around 40,000 SEAM doctorates graduating annually, China is currently close to the US in number, while still being well behind (ca. 37%) the number of STEM PhDs graduating from EU universities.

**Table 2 Number of STEM PhD graduated**

|      | China  | EU     | US     |
|------|--------|--------|--------|
| 2000 | 9,038  | NA     | 27,862 |
| 2005 | 20,269 | NA     | 34,468 |
| 2010 | 34,801 | NA     | 36,711 |
| 2015 | 40,963 | 60,223 | 44,521 |

Source: NSF (2018); CBS (2018), ESTAT (2018)

The expansion of China's higher education system offers only a partial view into the processes leading to the strengthening of China's scientific human capital base. A key part of this development has included the very large number of outbound students and scientists who travelled to the US, Europe and other countries to study and work (Simon and Cao, 2009; Jonkers, 2010). The outward mobility of Chinese students and scholars overseas began immediately after the Cultural Revolution in 1976. At that time, the Chinese leadership realized the negative impacts on the economy and defence of a "missing generation" of young and middle-aged, well-trained scientists, engineers, and other professionals on the nation's modernization in agriculture, industry, science and technology, and national defence. In addition to resuming formal higher education, the leadership put the training of high-quality personnel overseas higher on its agenda. As early as 1978, Deng Xiaoping, who would become China's paramount leader, proposed sending thousands and even tens of thousands students and researchers abroad as one of the important ways to raise the level of Chinese science and education. He did not perceive "brain drain" to be a problem, as long as ten percent of the dispatched would return.

The open-door policy (Zweig, 2002) also marked an important cultural shift which accompanied a more 'technical' or development policy driven shift. This cultural shift was started by the strong statements in support of S&T and scientists made by Deng Xiaoping at the first National Science Conference (1978). Not only were science and scientists respected again. Possibilities were also created to go and learn abroad, with the idea that they would bring this experience back to China. The process of opening up continued apace until the 1989 pro-democracy movement; thereafter, some restraints were added. In 1993, the twelve-word policy was adopted that again supported outflows of students and scholars. These twelve words refer to "support study overseas, encourage returns and guarantee freedom of (international) mobility" (*zhichi liuxue, guli huigou, laiqu ziyou*), with four more words, "allowing them to play a role" (*fahui zuoyong*) being added recently by Xi Jinping, the general secretary of the CCPCC and state president (for an analysis of the evolution of policy developments alongside a description of outbound and return flows, see Zweig and Chen (1995), Zweig (2002), Xiang (2003) and Cao (2008)).[iv]

Consequently, by 2017, a stock of 5.2 million Chinese had gone abroad and 3.1 million had returned; the return rate appears to be about 60.3%. Taking into consideration that 1.45 million were still students, 83.7% of those who had finished their overseas studies had returned. By comparison, by 2000, there had been only 340,000 and 130,000 overseas students and returnees. In terms of flow, in 2017, 608,400 Chinese went abroad and 480,900 returned, representing a return rate of 79% in the year. By comparison, in 2000, only 38,989 and 9,121 Chinese went abroad and returned respectively with a rate of return of 23.4% in the year.

Over time, some of these scientific émigrés returned to China. Indeed the two processes are seen as interlinked: the large investments in the higher education and research system, coupled with its qualitative upgrading, have made China attractive for returnees. Upon return, they were expected to help strengthen China's research and innovation system.

**Table 3 students going overseas and returning**

| Year | Number of students going overseas | Number of returnees | Cumulative number of students going overseas | Cumulative number of returnees |
|---|---|---|---|---|
| 2000 | 38,989 | 9,121 | 340,000 | 130,000 |
| 2005 | 118,515 | 34,987 | 933,400 | 232,900 |
| 2010 | 284,700 | 134,800 | 1,905,400 | 632,200 |
| 2017 | 608,300 | 480,900 | 5,194,900 | 3,132,000 |

*Source: compilation on the basis of data released by China's Ministry of Education.*

The bulk of the returnees earned Bachelor and Master level degrees in science, technology, engineering, and math (STEM). The return rate for Chinese with EU doctorates is thought to be higher than that from the United States. This is corroborated by the 'stay rates' found among the Chinese who have earned American doctorates, reported by the U.S. National Science Foundation. Between 2006 and 2016, US universities awarded 50,439 doctoral degrees to Chinese nationals (US NSF, 2018).[v] In 2015, 22% of the 464,000 foreign born S&E doctorate holders in the USA were Chinese. Between 2012 and 2015, the vast majority of U.S. science and engineering (S&E) doctorate recipients from China (83%) reported plans to stay in the United States, and approximately half of these individuals reported accepting firm offers for employment or postdoc research in the United States. By country of citizenship at time of degree, China, the country that is the source of more S&E doctorate recipients than any other foreign country, had the highest 5- and 10-year stay rates. For those who received their doctorates in 2005, the 10-year stay rate was 90%; for those who received their doctorates in 2010, the 5-year stay rate was 85%, while the 5-year and 10-year stay rates for all the S&E doctorates were both 70% in 2015 (US NSF, 2018). In the life sciences in 2007, there were 2,500 Chinese origin scholars serving as faculty members in US universities (Wang, 2007). In contrast to the US, there are no systematically collected statistics on the number of Chinese students, scientists and returnees in the EU, limiting the potential for comparative analysis. This is one of the gaps that this study aims to address.

**Table 4 Doctorates awarded at American universities to Chinese citizens: 2006–16**

|      | Doctorates | STEM doctorates |
|------|------------|-----------------|
| 2006 | 4,448      | 4,123           |
| 2010 | 3,744      | 3,457           |
| 2015 | 5,374      | 4,970           |

*Source: US NSF, 2018*

To counter the perceived "brain drain" and to further encourage return, China has implemented a series of programmes to attract back highly skilled overseas talent to strengthen its universities, research institutes and high-tech companies. As policy background to this paper, Table 5 provides an overview of these programmes as well as the number of researchers involved.

**Table 5 Programs Related to the Talent Attraction, Retaining and Utilization (by 2018)**

| Program | Agency in charge | Target of the program | Year initiated | Total affected number |
|---------|------------------|----------------------|----------------|----------------------|
| Hundred Talent Program | CAS | scientists under 45 years old (i) | 1994 | n.a. |
| National Science Fund for Distinguished Young Scholars | NSFC | academic leaders under 45 old; frontier sciences and technology (d) | 1994 | 3454 |
| Chunhui Program | MOE | Chinese expatriates for short-term services (i) | 1996 | n.a. |
| Cheung Kong / Changjiang Scholar Program | MOE | Endowed professorships for under 45 years old; extended to 55 years old in social sciences and humanities (i) | 1998 | 2948 |
| 111 Program | MOE & SAFEA | 1,000 foreign scholars from the top 100 universities and research institutions (i) | 2005 | n.a. |
| Thousand Talent Program | CLGCTW | 1,000 academics, corporate executives, and entrepreneurs under 55 years old to return from overseas | 2008 | n.a. |

| Young Thousand Talent Program | CLGCTW | academics under 40 years old with three+ years of post-doctoral research (i) | 2010 | 3535 |
|---|---|---|---|---|
| Science Fund for Emerging Distinguished Young Scholars | NSFC | researchers under 38 years old to work in academia (d) | 2011 | 2398 |
| Ten Thousand Talent Program | CLGCTW | To support high-end talent residing in China (d) | 2012 | 3454 |
| New Hundred Talent Program | CAS | Renewal of Hundred Talent Program (d & I) | 2014 | n.a. |
| Young Cheung Kong Scholar Program | MOE | Endowed professorships for young scholars at Chinese universities (d) | 2015 | 440 |

Notes: MOE – Ministry of Education; CAS – Chinese Academy of Sciences; NSFC – National Natural Science Foundation of China; SAFEA – State Administration of Foreign Expert Affairs; CLGCTW – Central Leading Group for the Coordination of Talent Work. (I = international focus; d=domestic focus). Source: author's review of grey literature and official sources.

A rough sum of the number of people attracted back through the various return programmes suggests that by 2018 the Chinese government had recruited back at least 16,000 scientists and high-tech entrepreneurs. Others may have returned following less well-known but presumably more extensive provincial or institutional programmes or on their own accord. Entrepreneurs and highly skilled employees beyond academia have also taken advantage of the opportunities offered by China's growing economy and high-tech development within the special economic zones (SEZs), introduced in the 1980s, as well as many high-tech parks in cities such as Beijing, Shanghai, Guangzhou, Shenzhen, Hangzhou, and beyond. Here policies granted benefits to returnees and their dependents to encourage highly skilled workers to locate in SEZs and high-tech parks (Liu & van Dongen, 2016). These benefits have since been expanded to other areas and, indeed, competition for talent has driven mobility policies at the regional level.

The impact of some of the return programmes is being studied. For example, Li & Tang (2019) examined the return and advancement patterns for Chang Jiang/Cheung Kong Scholars (n=1447) (see Table 5) and found that international mobility had a heterogeneous effect on career progression within China, with overseas experience slowing down the advancement of some late-phase careers and appeared to make little difference to early-stage careers, but local connections appear to be more important to advancement. Li, Miao and Yang (2015) analysed differences in the publication behaviour of Chang Jiang Scholars finding that those who returned to their alma mater collaborated less intra-institutionally and tended to publish higher impact publications. The results of the current study are also in line with a recent analysis of the Young Thousand Talent programme (Yang & Marini, 2019)

A full assessment of the relative success of the individual programmes described in Table 2 is beyond the scope of this paper, as it would require detailed personnel data on the successful and unsuccessful applicants to these respective programmes. This paper addresses the significance of the mobility of scientists between the Chinese as well as the United States' and European Science system. At present no comparable statistics exist of the number of overseas Chinese scientists working in the US and Europe, nor of the number of Chinese scientists who have returned to work in the Chinese science system following training and work abroad.

## 4. Methodology and Data

In the absence of reliable statistics, scientific mobility patterns have been studied using micro-data from surveys (e.g. Cruz-Castro et al, 2015; Franzoni et al, 2012; 2015; Scellato et al, 2015; Baruffaldi and Landoni, 2012) or collected from targeted samples of returnees (Jonkers & Tijssen, 2008; Jonkers and Cruz-Castro, 2013). Micro-data collected through surveys has well-known drawbacks. For example, given the need for anonymisation, it is often impossible to link micro-data to objective measures of productivity or collaboration. In the early 2000s, scholars started to explore the potential of bibliometric data to study scientific mobility and its effects (e.g. Laudel, 2003, Jonkers 2008).

A new bibliometric approach offers an alternative possibility to address both issues (building on Wagner et al, 2018; Sugimoto et al, 2017; OECD, 2017; Moed and Plume, 2012). An analysis of the micro-data on publications contained in Elsevier's Scopus database allows us to track researchers from the moment of first publication. It is thus possible to trace Chinese researchers who have first published in China and subsequently published in a different country. The number of researchers who started their publishing career in China, followed by publications in a host country, and who then continue to publish at the time of the analysis (i.e. contained in the Scopus database per 2019-01-29) are used as a proxy for mobility. This measure has limitations. In 2003, 75% of Chinese origin PhD holders working in the US had received their PhD degree from US universities (Cao, 2008). This share may be lower at present but it is expected to be substantial. Many of these researchers will not have published in journals indexed in Scopus before moving to the US and are therefore not captured in our analysis. However the measure does offer an important improvement/alternative over current surveys. As the same approach is followed for the US and EU, the statistics are comparable, thus addressing the problem that in many European countries no comparable data is available on scientists with a Chinese origin. The method also allows us to identify active scientists, excluding those overseas Chinese PhD holders who pursue a different career that does not involve publishing.

A proxy for returnees to China is calculated in a similar fashion. Of the scientists publishing with a Chinese address, a subgroup of authors is identified who had previously published with a US or EU address respectively. While this group of authors may include researchers with a US or EU origin, the vast majority are expected to be returned overseas Chinese scientists (Jin et al, 2007).

Publishing authors in the Scopus database (Schotten, el Aisati, Meester, Steiginga, & Ross, 2017) are defined by their author profile. Scopus is among the largest curated abstract and citation databases consisting of metadata of over 76 million scientific publications published since 1788 to date (Baas, Schotten, Plume, Côté, & Karimi, 2019). Author profiles in Scopus are a combination of curated and system-generated profiles. All author profiles are originally generated by an "author profiling" algorithm. This results in profiles that are optimized for precision (publications merged in a profile belong to one and the same person) over recall (all publications of the same person are merged into one profile). About 1.8 million profiles in Scopus have also undergone manual curation. The current reported precision and recall by Scopus is 98.1%, at an average recall of 94.4% (Baas, Schotten, Plume, Côté, & Karimi, 2019). The database captures links between authors and their affiliations on papers throughout the database historically. This allows for longitudinal researcher-mobility studies: allowing to track publications of the same researcher, their affiliations over time, using a comprehensive and curated database. This provides the basis of the mobility data presented in this paper.

In order to determine the flow of researchers, this study builds further on the methodology of the OECD (OECD, 2017, p. 71) by tracking author flows in Scopus using the first publication in each year and assigning authors fractionally to countries if that publication has multiple affiliation-countries for that author (for instance, if an author has a guest position, it can be practice to list multiple affiliations). Having a sequence of countries per author over time allows tracking of movement of that author. It also provides a point-in-time identification after which an author has moved to a tracked destination. For example, a researcher publishing in China first in 2005, moving abroad and publishing in the USA in 2007, and then returns to and publishes in China in 2014, is classified as a China returnee. We then track publications of this returnee after 2014. Similarly, if an author moves from the EU to China and back to the EU and back to China, the returnee contribution to international publications for China is measured from the first move to China onwards, and publications authored while not in China (back in the EU) are not counted.

Using this approach, it is possible to determine, for instance, the contribution of returnees from the USA to co-publications between China and the USA, and to map their relative impact against the total of international publications. Movements between two countries are registered in this study only if they are subsequent.. This means that if an author moves from the USA to EU and eventually to China, only the movements from the USA to the EU and from the EU to China are reported, and a movement from the USA to China is not taken into consideration. Contributions to publications after moving into a destination are measured at the time of the first publication at the destination.

In order to estimate stocks of overseas researchers and returnees, it should be taken into consideration that some researchers do not publish every year. To be able to count those authors even when they are not publishing in a given year, there are two different scenarios to take into account: (1) a researcher does not publish for a while and then publishes again; these authors are counted towards the last year available until they publish again, i.e. the missing years in between are filled using the last known data point. (2) a researcher that stops publishing and does not publish again (until today); for these an arbitrary threshold of 2 years is used in which they are still counted, until eventually they are classified after that period as "retired" and no longer counted. This is mainly relevant for the more recent years where it is important to be able to count stocks of researchers that haven't published in the last year, but for which it is not certain if they are retired or just temporarily inactive.

Access to micro-data on overseas Chinese scientists and returnees allows us to engage in a further analytical step that sheds insight on the contribution these scientists have made to the Chinese, EU and US science systems and on the relations between them. The first step is to assess the relative share of top 10% most highly cited papers from the different science systems co-authored by overseas Chinese and returnee authors. Apart from contributing to the human resource in science and technology (HRST) stocks in the US, EU and for returnees the HRST stocks in China, scientific mobility flows are also thought to play an important role in transnational knowledge transfer and to be closely related to international collaboration (Wagner, 2008). The scientific social capital, i.e. the scientific networks scientists had built up during their stay in a foreign country is expected to shape collaboration patterns at later stages in their career – including upon return (Jonkers, 2010; Jonkers & Tijssen, 2008; Jonkers and Cruz Castro, 2013). The second step is therefore to assess the share of international co-publications between the US and China as well as between the EU and China respectively. If co-publications are more prevalent in the direction of the former host system, then this substantiates the often referenced hypothesis that scientific social capital influences collaboration patterns. If we cannot show tendency towards co-publications, then this longer term effect of scientific mobility may not occur. International co-publications tend to receive more citations than the average paper published in each of the science systems (Royal Society, 2011). The participation of overseas Chinese scientists and returnees in co-publications with their EU and US counterparts respectively thus sheds light on the mechanisms underlying the sustained contribution of scientific mobility flows to the performance of advanced and emerging science systems.

5. **Results: assessing the size and impact of overseas Chinese scientists and returnees**

Using the approach described in the methodological section it is possible to make an estimated assessment of the stocks of people who returned to China, as well as recent mobility. Figure 2 and Table 6 present these data.

**Figure 2 Stocks of mobile Chinese scientists in China, the EU and US**

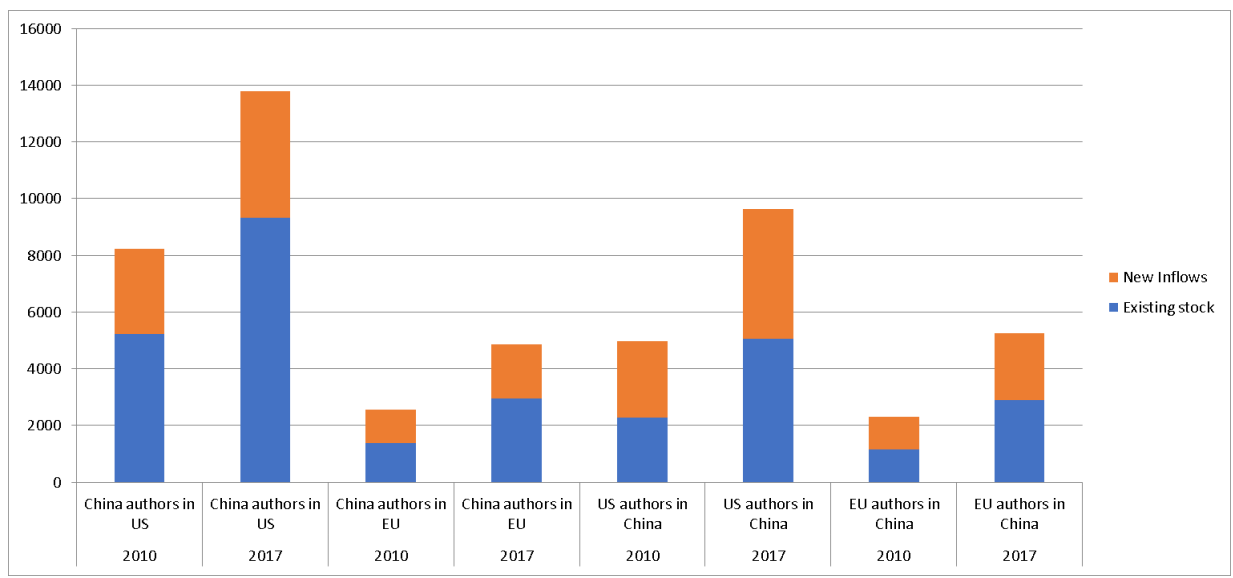

|  |  | Chinese authors now publishing in the US | Chinese authors now publishing in EU | US authors now publishing in China | EU authors now publishing in China |
|---|---|---|---|---|---|
| 2010 | Preceding stock (movement prior to 2010) | 5222 | 1386 | 2279 | 1163 |
|  | new movement | 3019 | 1175 | 2703 | 1141 |
| 2017 | Preceding stock (movement prior to 2017) | 9321 | 2957 | 5058 | 2889 |
|  | new movement | 4453 | 1905 | 4569 | 2371 |

Figure 2 shows estimates of the number of people who left China and now publish in the EU or the US, and the number of people who left the EU or the US and now publish in China. Column 1 in the table under the figure shows the number of scholars who had previously published in China and now publish in the US (5222) and the number of new authors in this category in 2010 (3019), and the same data for 2017, that is prior mobile scholars in the US at 9321 (including all previous years, including the 2010 data), and those new in 2017 (4453). The top section of the bars in Figure 2 shows the new inflows in 2010 or 2017; the 'existing stock' (prior mobility) is shown in the bottom part of the column. Those authors who began publishing in the US prior to 2010 and returned to China are shown in Table 6, column 3, at 2279, and those US authors newly publishing in China in 2010 (2703) and the same data for the EU in that year. The data showing those authoring first in the US and EU and moving to China (approx. 9,630 and approx. 5,260 respectively) primarily involves Chinese researchers who published in the US or the EU and then returned to China.

One observes from Figure 2 that between 2010 and 2017 an increasing number of Chinese researchers returned home from the US and EU, but the number of publishing Chinese researchers moving to the US and EU is higher than those returning home. In 2017 the number of researchers who had started their publication career in China and subsequently moved to the US is around three times higher (approx. 13,770) than the number who moved to Europe (approx. 4,860). In 2017, the number of new Chinese entrants to these regions had grown in both the US and the EU.

The number of Chinese researchers going to the EU has grown faster than those going to the US, albeit from a lower base. The number of new entrants in the US in 2017 is 2.3 times higher than the number of new entrants in the EU. The ratio between overseas Chinese and returnees provides us with an insight in the relative return rates from the EU and US. Whereas for every returnee from the US, 1.4 overseas Chinese scientists remain in the US; this ratio is 1 to 0.9 in the case of the EU.

Using the same approach it is also possible to estimate the relative impact which returnees and overseas Chinese have on the Chinese science system. These data also provide insight into the performance differences between the researchers that remain overseas and those who return. Figure 3 shows that the total population of returnees (ALL->China) published nearly 14% of China's publications in 2005. This share remains relatively stable over time at 13% in 2010 and 2017. (The supplement provides the entire time series). Figure 1 shows returnees from the US (US->CHN) at the considerably larger number (as noted above), who also published a larger number of indexed publications than Chinese researchers returning from Europe (EU->CHN). Returnees from the US are responsible for 5.1 % of all China's publications whereas returnees from Europe are responsible for 2.9 %. The publication productivity of overseas Chinese scientists in the US or EU are not counted as part of the Chinese output.

**Figure 3 Publications made by overseas Chinese and returnees relative to total output of China**

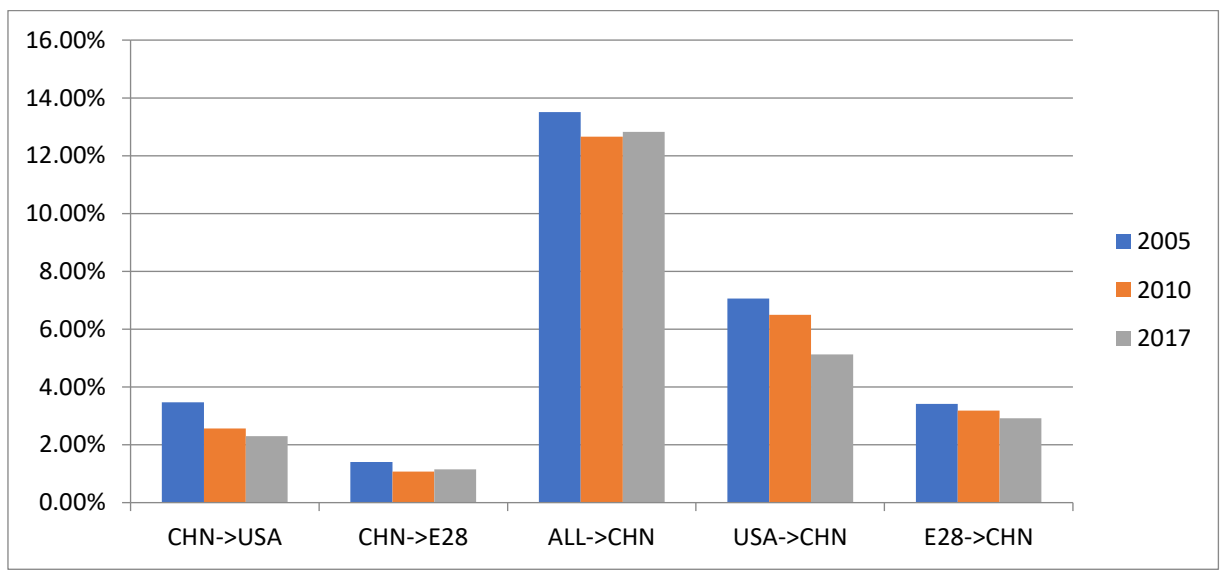

|      | CHN->USA | CHN->EU28 | ALL->CHN | USA->CHN | EU28->CHN |
|------|----------|-----------|----------|----------|-----------|
| 2005 | 3.5%     | 1.4%      | 13.5%    | 7.1%     | 3.4%      |
| 2010 | 2.6%     | 1.1%      | 12.7%    | 6.5%     | 3.2%      |
| 2017 | 2.3%     | 1.2%      | 12.8%    | 5.1%     | 2.9%      |

The column CHN->USA and CHN->EU28 depicts the productivity of this overseas Chinese population relative to the total Chinese output. For Chinese scholars in the US, productivity over-performance was 3.5% in 2005 and 2.3% in 2017. The over-performance of Chinese scholars in the EU was 1.4% in 2005 and 1.15% in 2017. In other words, Chinese researchers in the EU published 1.15 papers for every 100 papers made in China in 2017.

Returnees make a substantial contribution to the impact of Chinese science. Figure 4 shows, as expected, that the average impact of the work of overseas Chinese is considerably higher than that of those who have not moved abroad. Chinese scientists who remain in the US and EU meanwhile not only publish a relatively high and larger number of articles, the share of these publications which are among the top 10 % worldwide is at over 20% and 17% in the US and EU28 respectively, considerably higher even then that of the returnees, at 13% for returnees from both the US and EU28.

**Figure 4 The share of high impact (top 10%, fracFWCI) publications made by overseas Chinese scientists and returnees**

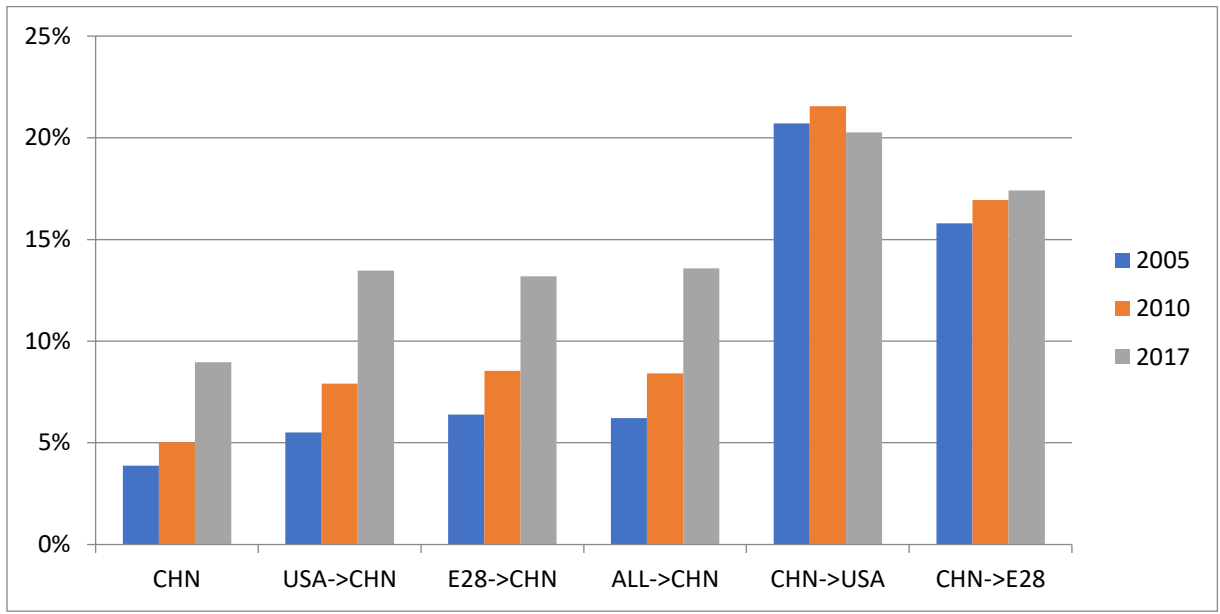

|      | CHN | USA->CHN | E28->CHN | ALL->CHN | CHN->USA | CHN->E28 |
|------|-----|----------|----------|----------|----------|----------|
| 2005 | 4%  | 6%       | 6%       | 6%       | 21%      | 16%      |
| 2010 | 5%  | 8%       | 9%       | 8%       | 22%      | 17%      |
| 2017 | 9%  | 13%      | 13%      | 14%      | 20%      | 17%      |

The share of China's publications that is among the top 10% most highly cited worldwide (adopting fractional counting) was 4% in 2005. This is well below the worldwide average of 10%. Over time, the percentage of highly cited publications by Chinese authors has been increasing, and can be expected to grow further. In 2010, the share of Chinese papers in the world's top 10% most highly cited (PP10) was 9%. As expected, the citation impact of papers by returnees is considerably above those who never left China. In 2005, this share was 2% higher, at 6% and in 2017 it was already 5% higher at 14%, i.e. 4 percentage points above the world average.

One of the factors related to stronger performance of returnees is their participation in internationally co-authored papers. Such papers are known in general to attract more citations and returned scholars use the connections built up during their stay abroad to continue publishing with foreign colleagues (Jonkers & Tijssen, 2008). Indeed if one considers the total share (fractional counting) of China's international co-publications published by researchers with foreign work experience, one observes that these shares are between 29% and 27% of China's total output of international co-publications, double the share of China-only publications. In other words, these researchers tend to publish a much higher share of papers as a result of international connections. The difference in the share of China's international co-publications by researchers who returned from the US, and those from the EU, is likely to be mainly a reflection of the different number of researchers returning from these parts of the world. What is striking is that a considerable share of the international co-publications is made by researchers who came to China from other parts of the world; in 2017, 27% of China's international papers were authored by returnees. This can be split into 16% US/EU and 11% from elsewhere. It suggests that flows of researchers from other East Asian countries, Canada, Australia, Singapore and the global south have had a considerable impact - also considering that a high share of these papers are among the top 10 most highly cited (PP10).

**Figure 5 International co-publications by overseas and returnees as a proportion of China's international co-publications**

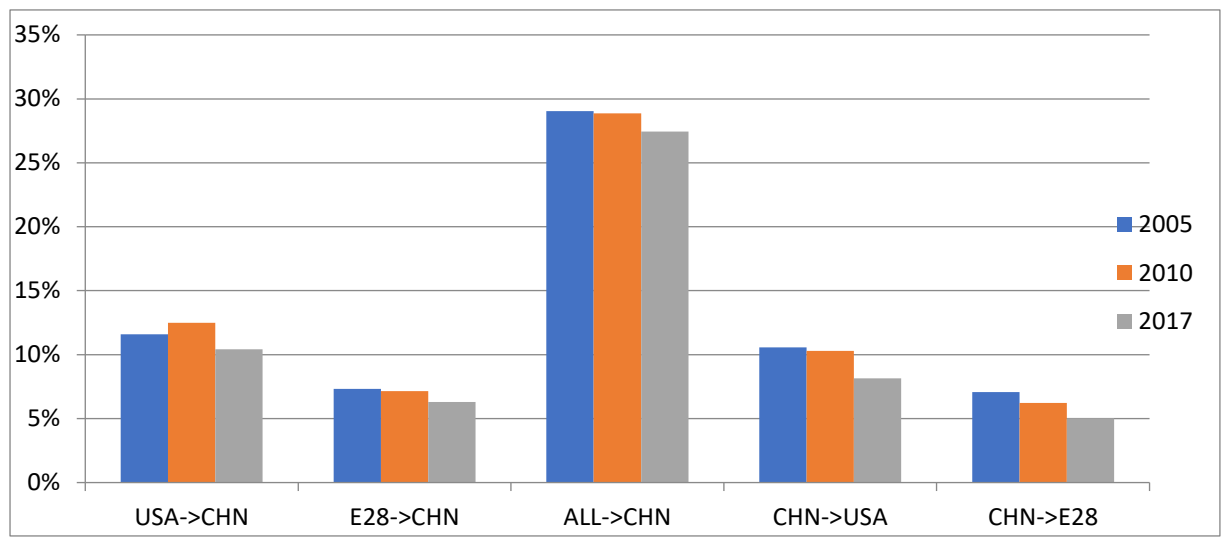

|      | USA->CHN | E28->CHN | ALL->CHN | CHN->USA | CHN->E28 |
|------|----------|----------|----------|----------|----------|
| 2005 | 12%      | 7%       | 29%      | 11%      | 7%       |
| 2010 | 12%      | 7%       | 29%      | 10%      | 6%       |
| 2017 | 10%      | 6%       | 27%      | 8%       | 5%       |

Chinese researchers who remain overseas also play an important role in forging ties between their host country and home system. As is clear from Figure 5, overseas Chinese scientists working in the US and EU are involved in 8% and 5% respectively of China's international co-publications in 2017. Figure 6 shows that researchers who moved to China after work experience in Europe predominantly co-publish with researchers in EU research systems, whereas researchers who moved to China from the US continue co-publishing predominantly with the US.

**Figure 6 the direction of co-publications by returnees**

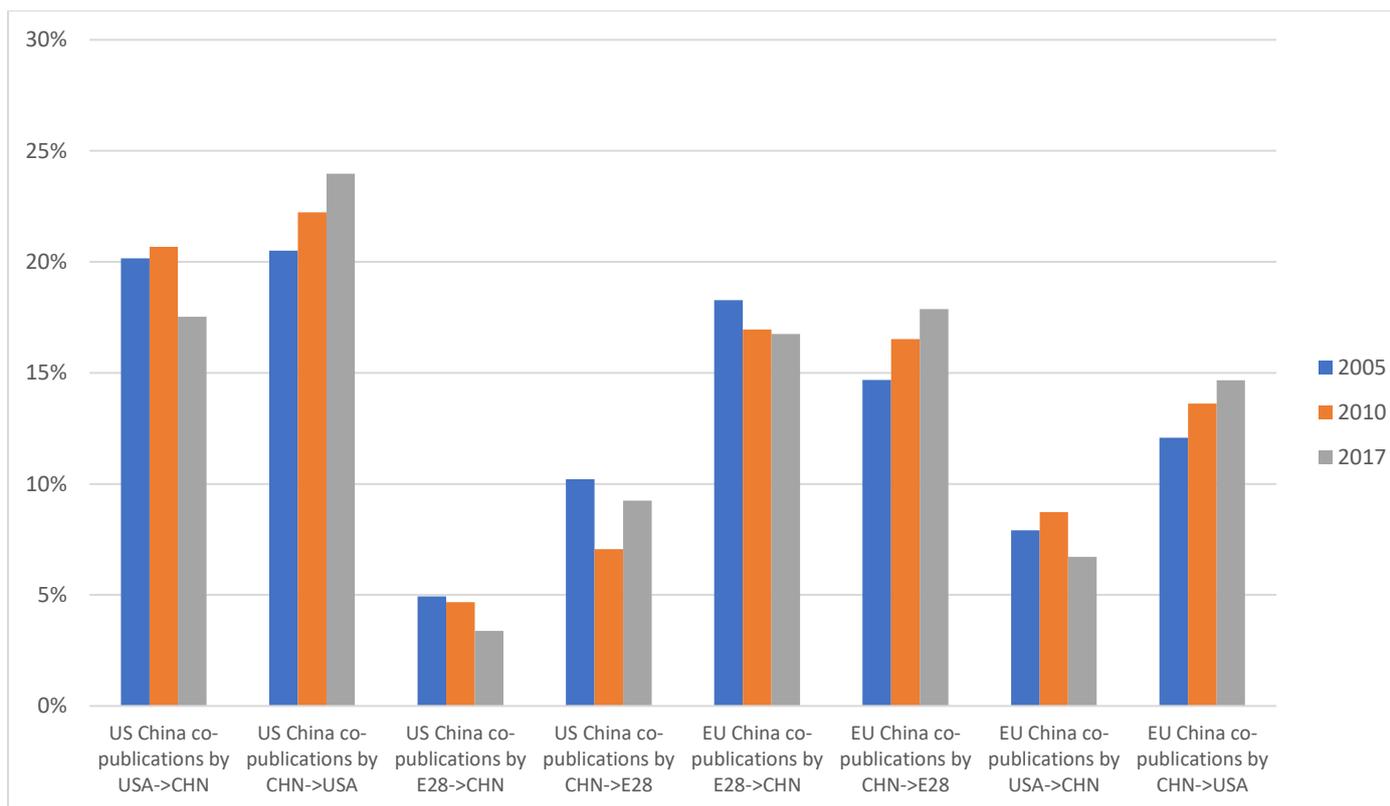

|   | US China co-publications by USA->CHN | US China co-publications by CHN->USA | US China co-publications by EU28->CHN | US China co-publications by CHN->EU28 | EU China co-publications by EU28->CHN | EU China co-publications by CHN->E28 | EU China co-publications by USA->CHN | EU China co-publications by CHN->USA |
|---|---|---|---|---|---|---|---|---|
| 2005 | 20% | 21% | 5% | 10% | 18% | 15% | 8% | 12% |
| 2010 | 21% | 22% | 5% | 7% | 17% | 17% | 9% | 14% |
| 2017 | 18% | 24% | 3% | 9% | 17% | 18% | 7% | 15% |

**Discussion and conclusion.**

The Chinese government's policy and programmes to build a world-class science system have borne fruit given the impressive rise of high impact science it produces. As shown in the data presented in this paper, the mobility (in and out) of foreign-bound students and scientists has had a major impact on the development of both the Chinese science system as well as the global science systems. Recognizing that the most highly talented of the overseas Chinese remained overseas, the Chinese government has actively sought to build and improve a national system to attract them back, and then has gone on a campaign to recruit them. While return migration programmes would in the best case scenario be able to attract back the best of the best, they might also claim success if they manage to attract people who substantially outperform those who have remained: given that their objective is to raise the performance of the domestic system. Even those who remain abroad appear to be an asset to China as co-authors. Countries facing large outbound flows of scientists may draw lessons from the Chinese experience.

One of the main challenges in assessing the impact of outbound and return flows of Chinese scientists to Europe and the US is the lack of comparable data on the population of overseas Chinese scientists. Data on returnees from these respective host systems is even less available. This article, rather than attempting to assess government return migration programmes, provides an insight in the relative dimensions of these stocks of scientists. As expected, we show that the share of overseas Chinese scientists in the US is considerably larger than that in the EU. Over time, flows of researchers from these destinations back to China have increased, although more so from the EU than the US. We also show that Chinese researchers who have returned from overseas both publish higher impact work, and continue to publish at the international levels and collaborate with the colleagues in the science systems in which they once participated.

As expected, Chinese scholars return to China in increasing numbers. We find their publications have a higher relative impact than domestic researchers. Further, we find that both the overseas Chinese and returned scientist population have contributed to the Chinese science system. It shows that over 12% of mainland China's total number of publications is published by people with overseas experience and, as explained in this paper, this number is likely to be a substantial underestimation. The share of high impact publications by scientists is considerably higher than that of their colleagues who remained in China throughout their scientific careers, as expected.

The impact of publications by overseas Chinese is higher than those of researchers in China (including returnees). This continues to be the case. In combination with a growing net outflow of Chinese researchers to the US and EU, this suggests that China may not yet attract back its 'best' expatriate scientists, or is perhaps not offering an environment for top-quality science (Cao, 2008). Many factors – not just talent – influence the quality and impact of publications. This includes the quality of the research environment, which though improving in China may still be less conducive to the high impact science China would like to conduct. Nevertheless, we show that overseas Chinese contribute to China through international collaborations. In the future China may succeed in capitalising further on the quite considerable contribution that overseas Chinese scientists make to the scientific effort of the USA and EU: in any case there appears to be some scope for this.

Part of the explanation for the relatively high impact of publications by returnees relative to their domestic counterparts is that they tend to engage more in co-publishing with foreign-based collaborators. This tendency is common among many mobile researchers and we found it to be true of China. Both returnees and overseas Chinese scientists play an important role in embedding China in international collaboration networks. The observation that returned Chinese scientists co-publish predominantly with researchers in their former host system is important for two reasons. First, it shows the importance of scientific social capital (Jonkers & Tijssen, 2008; Jonkers & Cruz-Castro, 2013): it is the contacts that people have built up in their host systems during their foreign work experience which continue to influence collaboration patterns upon return. It also shows the broader significance for any country of attracting foreign scientists. Mobile scientists play an important role in forging international collaboration ties, so a relative lack of attractiveness can influence long term integration with China and by extension the global science system.

In the analysis presented here, we have not distinguished between scientists on the basis of their length of stay and the stage of their career: the only criterion of foreign work experience was having published in a host system. This means returnees include junior scholars (such as the large number of doctoral students funded through the China Scholarship Council) as well as established ones who returned home after perhaps years away. This limitation should be kept in mind. Future research will seek to address this gap. As suggested in the return migration literature, this time factor can be important both for the accumulation of various forms of capital as well as for the successful reintegration upon return (King, 1986; Cassarino, 2004; Jonkers, 2010; Andújar, Cañibano & Fernandez-Zubieta, 2015). Cassarino (2004) classified different types of returnees as well as motivations to return. Whereas other methods are required to understand return migration decisions it may be possible to use the method described in this paper to perform further quantitative analysis of different types of returnees, by analysing e.g. those who return permanently or those who engage in multiple moves between their home and host system.

Outbound scientific mobility and return are shown in this paper to have an impact on both the host and the home system in terms of the production of high impact science and in shaping international collaborative ties. China is unique among late industrializers in the scale and scope of mobility as a strategy for development. The idea of advancing-through-learning is common to these countries, seen in Korea and Japan, but China is focused more on talent than on import substitution and imitation seen in other Asian nations. China's model has included an emphasis on basic science that was not a part of early actions of other 'Asian Tigers.' This feature—described here as mobility—may be a signature of China's rise as they have sought to benefit not only from the human scientific and technical human capital that these scientists constitute, but also from those benefits that accrue from greater integration in collaborative international networks (Wagner, 2018).

The large mobility flows and international collaboration patterns between China, the US and the EU have developed in a political context which allowed for these patterns of interactions to emerge in the decades since 1978. This political context might now be changing in particular with respect to the US and China. The Trump administration has singled out specifically China's Thousand Talent Program for scrutiny. There is concern in the academic community on both sides of the Pacific regarding the potential ramifications of these changes, although it is too early to tell whether they will have an impact on mobility, return and international collaboration.

**Bibliography**

Adams, J., (2012) 'Collaborations: The Rise of Research Networks', *Nature*, 490: 335–336.

Andújar, I., Cañibano, C., & Fernandez-Zubieta, A. (2015). 'International stays abroad, collaborations and the return of Spanish researchers', *Science, technology and society*, *20*(3), 322-348.

Appelbaum, R.P., Cao, C. Han, X. Parker, R., Simon, D. (2018). *Innovation in China: Challenging the Global Science and Technology System.* Cambridge, UK: Polity

Baas, J., Schotten, M., Plume, A., Côté, G., & Karimi, R. (2019). 'Scopus as a curated, high quality bibliometric data source for academic research in quantitative science studies', *Quantitative Science Studies*, forthcoming.

Baruffaldi, S. H., and Landoni, P. (2012). 'Return mobility and scientific productivity of researchers working abroad: the role of home country linkages'. *Research Policy* 41, 1655–1665. doi:10.1016/j.respol.2012.04.005

Borjas, GJ, (1994), 'The Economics of Migration, Journal of Economic Literature', *Journal of Economic Literature,* Vol. XXXII, pp 1667-1717

Borjas, GJ, Bratsberg, B., (1996), 'Who Leaves? Outmigration of the foreign born', *Review of Economics and Statistics*, 78, pp. 165-176

Bozeman, B., Dietz, J. S., and Gaughan, M. (2001). 'Scientific and technical human capital: an alternative model for research evaluation', *Int. J. Technol. Manag.* 22, 716–740. doi:10.1504/IJTM.2001.002988

Cao, C., (2008) 'China's brain drain and brain gain: Why government policies have failed to attract first-rate talent to return?', *Asian Population Studies*, 4 (3): 331-345.

Cao, C., Suttmeier, RP, (2017), 'Challenges of S&T System Reform in China', *Science*, 355 (6329): 1019-1021.

Cao, C., Suttmeier, RP., Simon, DF., (2006), 'China's 15-Year Science and Technology Plan', *Physics Today*, Vol. 59, No. 12 : 38–43..

Cassarino, J-P, (2004), 'Theorising Return Migration: The Conceptual Approach to Return Migrants Revisited'. International Journal on Multicultural Societies (IJMS),Vol. 6, No. 2, pp. 253 -279, 2004. Available at SSRN: https://ssrn.com/abstract=1730637

Cimini, G., Zaccaria, A., and Gabrielli, A. (2016). 'Investigating the interplay between fundamentals of national research systems: performance, investments and international collaborations.' *J. Informetrics* 10, 200–211.

CCPCC, (2010), The Central Committee of the Chinese Communist Party and the State Council of China, 'National Medium and Long-Term Plan for the Development of Talent (2010–2020)' (in Chinese), available online at http://www.gov.cn/jrzg/2010-06/06/content_1621777.htm (accessed on June 30, 2018).

Cruz-Castro L., Jonkers K., Sanz-Menéndez L. (2016) 'International Mobility of Spanish Doctorate Holders', In: Gokhberg L., Shmatko N., Auriol L. (eds) *The Science and Technology Labor Force*. Science, Technology and Innovation Studies. Springer, ISBN 978-3-319-27210-8

Edler, J., Fier, H., & Grimpe, C. (2011). 'International scientist mobility and the locus of knowledge and technology transfer', *Research Policy*, 40(6), 791-805. https://doi.org/10.1016/j.respol.2011.03.003

Franzoni, C., Scellato, G., and Stephan, P. (2012), 'Foreign-born scientists: mobility patterns for 16 countries', *Nat. Biotechnol.* 30, 1250–1253. doi:10.1038/nbt.2449

Franzoni, C., Scellato, G., and Stephan, P. (2015). 'International mobility of research scientists: lessons from GlobSci', in A. Geuna, A., (ed) *Global Mobility of Research Scientists—The Economics of Who Goes Where and Why*, Cambridge (US) Academic Press. ISBN: 9780128013960

Gaule, P., Piacentini, M., 2013, 'Chinese graduate students and US Scientific Productivity', *Review of Economics and Statistics*, 95 (2) p.698-701

Gibson, J., McKenzie, D, 2014, 'Scientific mobility and knowledge networks in high emigration countries: Evidence from the Pacific', *Research Policy*, 43(9), p. 1486-1495.

Scellato, G., Franzoni, C., Stephan, P, (2015) 'Migrant scientists and international networks', *Research Policy,* Volume 44 (1), Pages 108-120 https://doi.org/10.1016/j.respol.2014.07.014

Jin, BH., Rousseau, R., Suttmeier, RP, Cao, C., (2007). 'The Role of Ethnic Ties in International Collaboration: The Overseas Chinese Phenomenon.' Pp. 427–436 in *Proceedings of the ISSI*

*2007 (11th International Conference of the International Society for Scientometrics and Informetrics)*, edited by Daniel Torres–Salinas and Henk F. Moed. Madrid, Spain: Centre for Scientific Information and Documentation of the Spanish Research Council.

Jonkers, K. (2010). *Mobility, Migration and China's Scientific Research System*, Routledge China Series, Milton Park, UK: Routledge.

Jonkers, K., and Cruz-Castro, L. (2013), 'International mobility, research collaboration and productivity of Argentinean life scientists', *Research Policy* 42, 1366–1377. doi:10.1016/j.respol.2013.05.005

Jonkers, K., and Tijssen, R. (2008). 'Chinese researchers returning home: impacts of international mobility on research collaboration and scientific productivity'. *Scientometrics,* 77, 309–333. doi:10.1007/s11192-007-1971-x

Jonkers, K., 2008, 'A Comparative Study of Return Migration Policies Targeting the Highly Skilled in Four Major Sending Countries', MIREM-AR; 2008/05; Robert Schuman Centre for Advanced Studies, Florence, Italy http://hdl.handle.net/1814/9454

King, R., ed. 1986. '*Return Migration and Regional Economic Problems*'. London: Croom Helm.

Laudel, G. 2003, 'Studying the brain drain: Can bibliometric methods help?', *Scientometrics,* 57 **:** 215–237.

Leydesdorff, L., Wagner, C. S., & Bornmann, L. (2014). 'The European Union, China, and the United States in the top-1% and top-10% layers of most-frequently cited publications: competition and collaborations', *Journal of Informetrics*, *8*(3), 606-617

Li, F., Tang, L. (2019), 'When international mobility meets local connections: Evidence from China', *Science and Public Policy*, Doi: 10.1093/scipol/scz004

Li, F., Miao, YJ, Yang, CC (2015). 'How do alumni faculty behave in research collaboration? An analysis of Chang Jiang Scholars in China', *Research Policy,* 44(2), 438-450, https://doi.org/10.1016/j.respol.2014.09.002

Liu, H., van Dongen, E. (2016). 'China's Diaspora Policies as a New Mode of Transnational Governance,' *Journal of Contemporary China* 25:102, 805-821. DOI: 10.1080/10670564.2016.1184894

Meyer, J-B., (2001), 'Network Approach versus Brain Drain, Lessons from the Diaspora', *International Migration*, 39(5): 91-110

Moed, H. F., and Plume, A. (2013). 'Studying scientific migration in Scopus'. *Scientometrics* 94, 929–942. doi:10.1007/s11192-012-0783-9

National Bureau of Statistics of China, 'Statistical Communiqué of the People's Republic of China on National Economic and Social Development in 2017 (February 28, 2018)' (in Chinese), available online at http://www.stats.gov.cn/tjsj/zxfb/201802/t20180228_1585631.html (accessed on June 30, 2018).

National Science Foundation, National Center for Science and Engineering Statistics, *National Patterns of R&D Resources: 2015–2016 Data Update*, NSF 18–309 (Alexandria, VA: U.S. National Science Foundation, May 2018 [revised June 2018]).

National Science Board (2018), *Science and Engineering Indicators 2018 Digest*. NSB-2018-2. Alexandria, VA: National Science Foundation. Available at https://www.nsf.gov/statistics/digest/.

NBS (2018), *China statistical yearbook of Science and Technology 2017*, Beijing: China Statistics Press

OECD (2010), *OECD Science, Technology and Industry Outlook 2010*, OECD Publishing, Paris, https://doi.org/10.1787/sti_outlook-2010-en.

OECD (2017), *OECD Science, Technology and Industry Scoreboard 2017: The digital transformation*, OECD Publishing, Paris, https://doi.org/10.1787/9789264268821-en.

Preziosi, N., Fako, P., Hristov, H., Jonkers, K., Goenaga, X. (eds) Amoroso, S., Alves Dias, P., Annoni, A., Asensio Bermejo, JM, Blagoeva, D., Bellia, M., De Prato, G., Dosso, M., Fako, P., Fiorini, A., Georgakaki, A., Gkotsis, P., Goenaga, X., Hristov, H., Jaeger-Waldau, A., Jonkers, K., Lewis, A., Marmier, A., Marschinski, R., Martinez Turegano, D., Munoz Pineiro, A., Nardo, M., Ndacyayisenga, N., Pasimeni, F., Preziosi, N., Rancan, M., Rueda Cantuche, JM, Rondinella, V., Tanarro Colodron, J., Elsnig, T., Thiel, C., Testa, G., Tuebke, A., Travagnin, M., Van den Eede, G., Vazquez Hernandez, C., Vezzani, A., Wastin, F. (2019)) '*China, challenges and prospects from an industrial and innovation powerhouse*', EUR 29737 EN, Publications Office, Luxembourg, 2019, ISBN 978-92-76-02997-7, doi:10.2760/445820, JRC116516.

Purkayastha, A., Palmaro, E., Falk-Krzesinski, H., & Baas, J. (2019). 'Comparison of two article-level, field-independent citation metrics: Field-Weighted Citation Impact (FWCI) and Relative Citation Ratio (RCR)' *Journal of Informetrics, 13*(2), 635-642. doi:10.1016/j.joi.2019.03.012

Royal Society, (2011) *Knowledge, Networks and Nations: Global Scientific Collaboration in the 21st Century. London: The Royal Society,*

Scellato, G, Franzoni, C., Stephan, P., (2015) 'Migrant scientists and international networks', *Research Policy*, Volume 44, Issue 1, Pages 108-120

Schotten, M., el Aisati, M., Meester, W., Steiginga, S., & Ross, C. (2017). 'A Brief History of Scopus: The World's Largest Abstract and Citation Database of Scientific Literature', In F.

Cantu-Ortiz, *Research Analytics. Boosting University Productivity and Competitiveness through Scientometrics, Auerbach publications,* ISBN 9781498785426

Simon, DF., Cao, C., (2009) *China's Emerging Technological Edge: Accessing the Role of High-End Talent*, New York: Cambridge University Press.

Simon, DF., Cao, C., (2011) 'Human Resources: National Talent Safari,' *China Economic Quarterly*, 15–19.

Sugimoto, C. R., Robinson-Garcia, N., Murray, D. S., Yegros-Yegros, A., Costas, R., and Larivière, V. (2017). 'Scientists have most impact when they're free to move', *Nature* 550, 29. doi:10.1038/550029a

Suttmeier, RP, Cao, C, Simon, DF., 2006, '"Knowledge Innovation" and the Chinese Academy of Sciences', *Science*, 312, 58-59

Sandström, U., Van den Besselaar, P., 2018, 'Funding, evaluation, and the performance of national research systems', *Journal of Informetrics*, 12/1, pp 365-384

Van Holm, E. J., Wu, Y., & Welch, E. W. (2018). 'Comparing the collaboration networks and productivity of China-born and US-born academic scientists'. *Science and Public Policy*. doi: 10.1093/scipol/scy060

Van Noorden, R., (2014) 'China Tops Europe in R&D Intensity', *Nature*, 505: 144–145;

Wagner, C.S. (2018) *The Collaborative Era in Science: Governing the Network*. London: Palgrave.

Wagner, C.S., and Jonkers, K. (2017), 'Open countries have strong science', *Nature* 550, 7674. doi:10.1038/550032a

Wagner, C.S. (2008). *The New Invisible College: Science for Development.* Brookings Press.

Wagner, C.S., Bornmann, L., & Leydesdorff, L. (2015), Recent developments in China–US in science, *Minerva*, *53*(3), 199-214.

Wang DH, (2007) 'Biologist Shi Yigong: The Twenty-first Century Is the Century of the Life Sciences' (in Chinese), *Science Times*, August 7, 2007, available online at http://news.sciencenet.cn/sbhtmlnews/200786234744428186299.html (accessed on June 30, 2018).

Wang, HY, (2011) 'China's New Talent Strategy: Impact on China's Development and Its Global Exchange', *SAIS Review*, Vol. 31, No. 2 (Summer–Fall 2011): 49–64

Wang, Q., (2017) 'Programmed to Fulfill Global Ambitions', *Nature*, Vol. 545 (May 25, 2017): S53.

Wang, QH, Wang, Q., Liu, NC., (2011) 'Building World-Class Universities in China: Shanghai Jiao Tong University', in Philip G. Altbach and Jamil Salmi (eds.), *The Road to Academic Excellence: The Making of World-Class Research Universities* (Washington, DC: World Bank, 2011), 33–62

Yang LL, Marini, G., (2019) 'Research productivity of Chinese young thousand talents', *International Higher Education*, No 97, DOI: https://doi.org/10.6017/ihe.2019.97.10944

Xiao, B., 2003, Emigration from China: A Sending Country Perspective, *International Migration*, 41 (3): 21-48

Zhang GL, (2018) 'Are 'Double First-Class' Universities Divided into Five Echelons?' (in Chinese), *Science and Technology Daily*, February 25, 2018, available online at http://news.sciencenet.cn/htmlnews/2018/2/403605.shtm (June 30, 2018).

Zhang, H., Patton, D., Kenney, M., (2013) 'Building global-class universities: Assessing the impact of the 985 Project', *Research Policy* Volume 42, Issue 3, Pages 765-775

Simon, DF, Cao, C., (2009) *China's Emerging Technological Edge: Assessing the Role of High-End Talent*, Cambridge: Cambridge University Press

Zhou, P., & Leydesdorff, L. (2006). 'The emergence of China as a leading nation in science', *Research policy*, *35*(1), 83-104.

Zweig, D., (2002) *Internationalizing China: Domestic Interests and Global Linkages*, Ithaca, NJ: Cornell University Press, pp. 161–210.

Zweig, D., Chen, CG, (1995) *China's Brain Drain to the United States: Views of Overseas Chinese Students and Scholars in the 1990s*, Berkeley, CA: Institute of East Asian Studies, University of California.

---

[i] Source: Scopus; extracted and indexed dataset at 2019-01-29.

[ii] "The CCPCC issued the Opinions on Deepening the Reform of the Institutional Mechanism for Talent Development" (in Chinese), available online at http://www.xinhuanet.com/politics/2016-03/21/c_1118398308.htm (accessed June 30, 2018).

[iii] In Chinese statistics information is collected on doctoral students majoring in science, engineering, agriculture, and medicine (SEAM). This classification is similar to the disciplines of science, technology, engineering, and mathematics (STEM) referred to in the context of other counties.

[iv] As an indication of the importance given to this issue at the highest level of government, Xi Jinping also praised Chinese studying overseas to be "the valuable treasure of the party and the Chinese people as well as the effective strength to realize the great rejuvenation of the Chinese nation." "President Xi Jinping's Speech at the 100[th] Anniversary of the Founding of

the Western Returned Scholars Association" (in Chinese), Available online at http://www.xinhuanet.com/politics/2013-10/21/c_117808372.htm (accessed September 28, 2019).

[v] National Center for Science and Engineering Statistics, Directorate for Social, Behavioral and Economic Sciences (comp.), *2016 Doctorate Recipients from U.S. Universities*, NSF 18–304 (Alexandria, VA: U.S. National Science Foundation, March 2018).  Table 26. Doctorates awarded for 10 largest countries of origin of temporary visa holders earning doctorates at U.S. colleges and universities, by country or economy of citizenship and field: 2006–16.